\documentclass[preprint,superscriptaddress,amsmath,amssymb,aip,jcp,floatfix]{revtex4-1}
\usepackage[utf8]{inputenc}
\usepackage{graphicx}
\usepackage{enumitem}
\usepackage{gensymb}
\usepackage{amsmath}
\usepackage{amsthm}
\usepackage{amssymb}
\usepackage{dcolumn}
\usepackage{color}
\usepackage{bbm}

\usepackage[normalem]{ulem}

\usepackage{multirow}
\usepackage{algorithm}
\usepackage{algpseudocode}
\usepackage{parskip}
\usepackage{makecell}

\begin{document}

\title{Graphic characterization and clustering configuration descriptors of determinant space for molecules}

\author{Lei Sun}
\thanks{These authors contributed equally to this work}
\affiliation{School of Physics, Peking University, Beijing 100871, P. R. China}

\author{Zixi Zhang}
\thanks{These authors contributed equally to this work}
\affiliation{School of Physics, Peking University, Beijing 100871, P. R. China}

\author{Tonghuan Jiang}
\email{jiangth1997@pku.edu.cn}
\affiliation{School of Physics, Peking University, Beijing 100871, P. R. China}

\author{Yilin Chen}
\affiliation{International Center for Quantum Materials, School of Physics, Peking University, Beijing 100871, P. R. China}

\author{Ji Chen}
\email{ji.chen@pku.edu.cn}
\affiliation{School of Physics, Peking University, Beijing 100871, P. R. China}
\affiliation{
 Collaborative Innovation Center of Quantum Matter, Beijing 100871, P. R. China
}
\affiliation{Interdisciplinary Institute of Light-Element Quantum Materials and Research Center for Light-Element Advanced Materials,Peking University, Beijing 100871, P. R. China
}
\affiliation{Frontiers
Science Center for Nano-Optoelectronics, Peking University, Beijing 100871, P. R. China}

\date{\today}

\begin{abstract}

Quantum Monte Carlo approaches based on the stochastic sampling of the determinant space have evolved to be powerful methods to compute the electronic states of molecules. These methods not only calculate the correlation energy at an unprecedented accuracy but also provides insightful information on the electronic structure of computed states, e.g. the population, connection, and clustering of determinants, which have not been fully explored. In this work, we devise a configuration graph for visualizing the determinant space, revealing the nature of the molecule's electronic structure. In addition, we propose two analytical descriptors to quantify the extent of configuration clustering of multi-determinant wave functions. 
The graph and descriptors provide us with a fresh perspective of the electronic structure of molecules and can assist the further development of configuration interaction based electronic structure methods.

\end{abstract}

\maketitle

\section{Introduction}

The exact wave function of realistic systems, e.g. large molecules, described by Schr{\"o}dinger's equation is generally out of reach because it is NP-hard to compute and hence approximations are often used \cite{PopleNobel,KohnNobel}. 
In the state-of-the-art wave function theory for the Fermionic system, one computational approach commonly adopted is based on expanding the many-body wave function by Slater determinants of single-particle orbitals, which are ultimately expanded in a given basis set. 
Theoretically, if one can include all possible determinant configurations unbiasedly, i.e. include the full configuration interaction (FCI), then the solution represents the exact solution in the given basis set.
In previous studies, the priority is mostly put on improving accuracy and efficiency, namely getting better energy calculation with lower computational cost,
but analyses of the wave function, i.e. configuration population in the Hilbert space, are relatively inadequate.
Such information is key to further developing efficient deterministic and stochastic wave function algorithms such as configuration interaction (CI), coupled cluster (CC), and many-body perturbation methods \cite{huron_iterative_1973,ivanic_identification_2001,knowles_compressive_2015,schriber_communication_2016,liu_ici_2016,holmes_heat-bath_2016,tubman_deterministic_2016,sharma_semistochastic_2017,deustua_communication_2018}.
Diagnostics of multi-reference characters have been proposed in various forms of electronic structure calculations.
However, some diagnostics may not capture the complete information of the multi-determinant wave function, e.g. counting the population of configurations but neglecting their connections.
In addition, existing diagnostics are often based on calculations with different approximations, e.g. density functional theory, multi-reference self-consistent field and coupled cluster \cite{duan_data_2020}, resulting in the discrepancies between various diagnostics.

Introduced by Booth et al. in 2009\cite{Booth_Thom_Alavi_2009}, full configuration interaction quantum Monte Carlo (FCIQMC) is one of the powerful algorithms in wave function theory, which employs an efficient stochastic sampling method to explore the full space of determinant configurations.
FCIQMC provides a promising means to tackle correlated electronic systems,
and the past decade has witnessed a rapid development of FCIQMC algorithm \cite{Cleland_Booth_Alavi_2010,Blunt_Smart_Booth_Alavi_2015,Ghanem_adapshift_2019}, as well as the range of applications \cite{booth_towards_2013,guther_neci_2020}.
Currently, analysis of FCIQMC simulations is limited to arranging the determinants according to its population in a one-dimensional histogram \cite{Chen_TiO2v_2020,jiang_full_2021}. 
For most systems, such an analysis is enough to tell whether the system is single-determinant or multi-determinant, but it is still not sufficient to understand the true nature of the determinant space.
Therefore, it is desirable to develop new analysis methods for FCIQMC simulations to provide deeper insights into the electronic structure of molecules.

In this work, we propose a new scheme of analysis based on a configuration graph using not only their populations but also the Hamiltonian matrix elements that represent the ``interaction'' between determinants.
The analysis is presented by a graph of nodes distributed in a two-dimensional manner.
We further propose clustering configuration descriptors to characterize the nature of determinant space, which includes but is not limited to the extent of multi-determinant.
In the following sections, we present details about FCIQMC simulation, which is the main methodology practiced in this work, the configuration graph, and the descriptors.
Based on these analyses, we further discuss the ground state of several molecular systems 
including $\text{H}_4$, $\text{B}_2$, $\text{C}_2$ and $\text{N}_2$, where distinct features are illustrated for molecules upon the variation of atomic geometry.
In addition, we show that the descriptors proposed may be used to further improve the existing electronic structure methods.

\section{Methods}
\subsection{FCIQMC}
The theoretical foundation of FCIQMC is the full configuration interaction (FCI) method, in which the wave function of a many-electron system can be written as

\begin{equation}
  \Psi=\sum_i C_i D_i, \qquad 
  D_i=\frac{1}{\sqrt{N!}}\text{det}
  \left|\begin{array}{ccccc} 
      \psi_{i1}(x_{1}) &    \psi_{i2}(x_{1})    & ... & \psi_{iN}(x_{1}) \\ 
      \psi_{i1}(x_{2}) &    \psi_{i2}(x_{2})   & ... & \psi_{iN}(x_{2})\\ 
      \vdots  & \vdots & & \vdots \\
      \psi_{i1}(x_{N}) & \psi_{i2}(x_{N}) & ... & \psi_{iN}(x_{N})
  \end{array}\right| 
\end{equation}

where \(D_i\) is the i-th configuration formed by a Slater determinant of molecular orbitals, 
\(C_i\) is the corresponding expansion coefficient, and N is the number of electrons.
If we start with the Hartree-Fock state, represented by one determinant formed with only occupied molecular orbitals, then one can interpret FCI as a method considering all configurations excited from the Hartree-Fock state, namely including all the electron correlations neglected in the Hartree-Fock wave function.

FCIQMC employs a stochastic algorithm to solve the FCI wave function. The deterministic calculation on the FCI wave function is too challenging \cite{Booth_Thom_Alavi_2009}.
The FCIQMC algorithm follows the Schr{\"o}dinger equation and its imaginary time evolution, which resembles the diffusion of classical particles. 
The evolution of wave function is discretized into many steps of propagation of Monte Carlo walkers.
The master equation is written as,
\begin{equation}
  C_i(\tau+\delta\tau)=(1-\delta\tau(H_{ii}-S))C_i(\tau)
  -\delta\tau\sum_{j\neq i}H_{ij}C_j(\tau)
\end{equation}
where $H$ is the Hamiltonian, $S$ is a parameter to be adjusted during the evolution for population control. 
Furthermore, each Monte Carlo step can be decomposed into three processes:
(i) spawning, where every configuration spawns a number of new walkers at another configuration, described by \(-\delta\tau\sum_{j\neq i}H_{ij}C_j(\tau)\);
(ii) cloning/death, where walkers of each configuration clone or die, described by \((1-\delta\tau(H_{ii}-S))C_i(\tau)\).
(iii) annihilation, where walkers with different signs on each configuration cancel each other so that only one type of walkers remains. 
The Monte Carlo simulation converges to the ground state of many-body electronic wave function as the imaginary time evolution proceeds \(\tau\rightarrow \infty\).

Our FCIQMC calculations are performed with the NECI package. \cite{guther_neci_2020}.
The cc-pVTZ basis set was used in our calculations on $\text{H}_4$, $\text{B}_2$, $\text{C}_2$ and $\text{N}_2$, respectively.
The molecular orbitals for the subsequent FCIQMC calculations were obtained with restricted Hartree-Fock, performed with the PySCF package \cite{PySCF}. 
The initiator (i-FCIQMC)\cite{Cleland_Booth_Alavi_2010} and adaptive shift approach (as-FCIQMC) \cite{Ghanem_adapshift_2019,ghanem_adaptive_2020} were used in all FCIQMC calculations, with initiator threshold $n_a=3$. 
The time-step was updated using the TAU-SEARCH facility of NECI.   
The semi-stochastic method \cite{petruzielo_semistochastic_2012, Blunt_Smart_Booth_Alavi_2015} was used, and the size of the deterministic space was set to 100. 
A trial wave function was used to obtain the projected energy estimate, and the size of the trial space was set to 100.

\subsection{Configuration graph}

\begin{figure}[h]
  \centering
  \includegraphics[width=16.0cm]{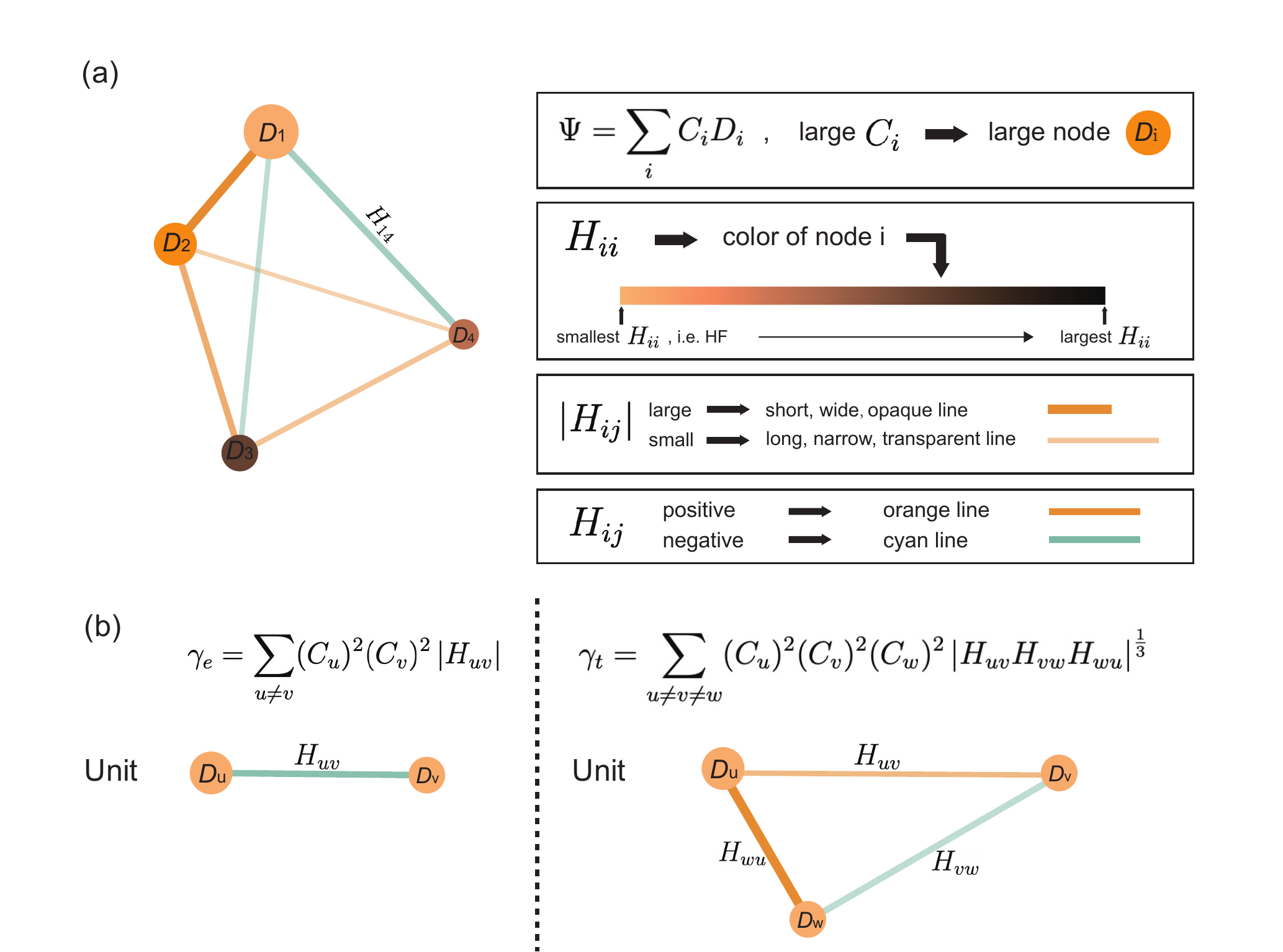}
  \caption{\textbf{Schematic diagram for the configuration graph and the clustering configuration descriptors}. (a) Description of the components of the graph. (b) Definitions and units for the $\gamma_e$ and the $\gamma_t$ descriptor based on edges and triangles, respectively.
  }
  \label{sketch}
\end{figure}

We design a configuration graph of the determinant space. The schematic diagram is shown in FIG.\ref{sketch}(a). The main elements of the graph are nodes with different sizes and colors, which are connected by lines varying in color and width. 
Each node represents a configuration $D_i$ and its size is proportional to the corresponding CI coefficient from FCIQMC simulation. 
The color of a node is related to the energy of the configuration itself, i.e., the diagonal elements $H_{ii}$ of the Hamiltonian matrix. 
Higher energy leads to darker nodes.
The length of lines is negatively correlated to the absolute value of the Hamiltonian matrix element between two configurations, i.e. the absolute value of off-diagonal elements $\left| H_{ij} \right|$. The width and transparency of lines are positively correlated to $\left| H_{ij} \right|$.
The color of lines is determined by the sign of $H_{ij}$, which takes orange if positive and cyan if negative.
The distance between two nodes is negatively correlated to $\left| H_{ij} \right|$. 
Hence, the configurations with greater Hamiltonian matrix elements tend to cluster together. 
In FCIQMC simulation a greater $\left| H_{ij} \right|$ means a greater chance of walkers spreading between the corresponding configurations. 

This graph provides a direct visualization of the FCI determinant space and shows the relationship between configurations.
A graph with nodes averagely scattered and connected by weak lines means the Hamiltonian matrix is highly diagonal and the walkers barely spread.
The leading configuration would have a CI coefficient of approximately 1, while all the other determinants are rarely populated. 
In this case, the exact FCI wave function would be highly similar to the Hartree-Fock wave function.
A similar situation occurs when the wave function is computed using single-determinant density functional theory.
For particular systems, we can see the gathering of nodes, where nodes are much closer to each other and have wider connected lines.
We call such a group a “cluster”.
In such a scenario, walkers could spread fast within a cluster and slowly between clusters, and the system would be far different from the single-determinant wave function.

It is worth noting that each pair of nodes has independent $H_{ij}$ and if we intend to draw a two-dimensional graph the distance between each pair is entangled and can not be determined separately. In this work, we use the two-dimensional (2D) Kamada-Kawai layout\cite{kamada_algorithm_1989} to get the position of nodes, which leads to an optimal determination of node separations following the value of $H_{ij}$.
A cross validation using the four-dimensional Kamada-Kawai layout in conjunction with the t-distributed stochastic neighbor embedding (t-SNE) algorithm \cite{t-SNE} shows that the main feature of the graph is correctly characterized by the 2D Kamada-Kawai layout.

\subsection{Clustering configuration descriptor}

In FCIQMC, the possibility of walkers spawning among configurations is proportional to Hamiltonian matrix elements. 
The configuration graph is a direct indicator of the possible dynamic behavior of walkers.
The distance between nodes is negatively correlated with the Hamiltonian matrix element, the farther the nodes are separated, the smaller the spawning probability is.
If the configurations form clusters, walkers would have a greater chance to spawn within the cluster.
Such features reflect some extent the ``interaction'' between different configurations and are likely to be related to the performance of traditional methods such as configuration interaction and coupled cluster.

The concept of the clustering coefficient in graph theory is designed to describe the degree of the gathering of nodes, and there are multiple ways to define it. 
Here we propose two descriptors as defined in Eq. \ref{descriptor ce def} and Eq. \ref{eq:definition_C1}.

\begin{align}
    & \gamma_e=\sum_{u\ne v}(C_u)^2 (C_v)^2\left| H_{uv}\right|
    \label{descriptor ce def}
\end{align}

\begin{align}
    & \gamma_{t} = \sum_{u\ne v\ne w} (C_u)^2(C_v)^2(C_w)^2\left| H_{uv}H_{vw}H_{wu}\right|^{\frac13} \label{eq:definition_C1}
\end{align}

$C_u$ is the CI coefficient at node $u$, and $H_{uv}$ is the Hamiltonian matrix element of node $u$ and node $v$.
The introduction of the CI coefficient guarantees the convergence of descriptors as a function of the number of configurations included.
$\gamma_e$ characterizes a pair of nodes in the graph, whereas 
$\gamma_t$ characterizes the triangles as shown in FIG. \ref{sketch}(b). 
The form $H_{uv} H_{vw} H_{wu}$ means the summation is based on triangles and each term represents an edge of the triangle composed of node $u$, $v$ and $w$.
When several nodes in a graph form clusters have large CI coefficients and are connected by large $H_{uv}$, descriptors $\gamma_e$ and $\gamma_t$ are large, indicating a multi-determinant system.
The two descriptors are both tensorially variant, and their values depend on the determinant space spanned by a given set of orbitals.

\section{Results}

\subsection{Visualization with configuration graph}

\begin{figure}[h]
  \centering
  \includegraphics[width=16.0cm]{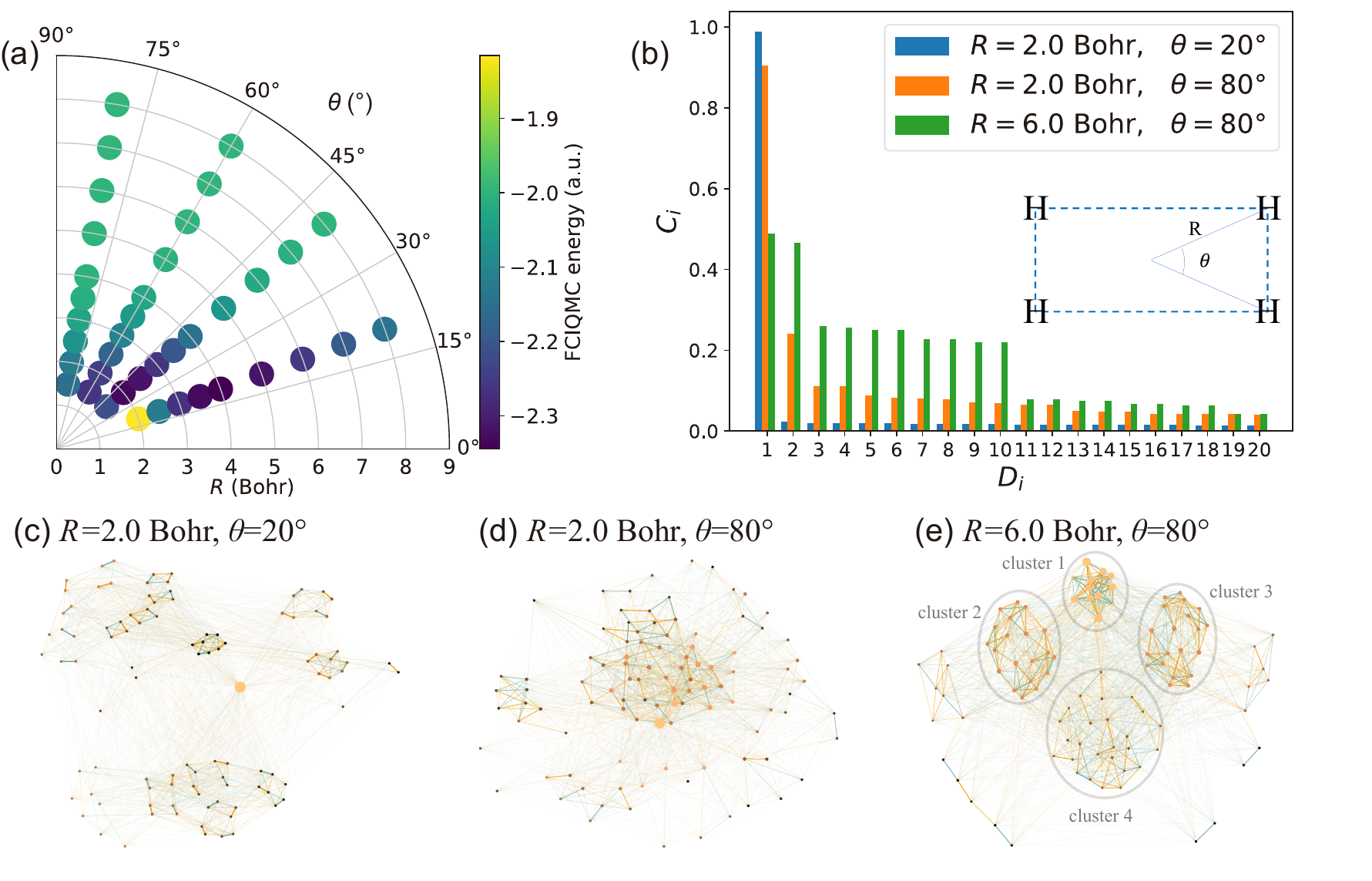}
  \caption{\textbf{FCIQMC calculation and configuration graphs of $\text{H}_4$.} (a) FCIQMC energy of $\text{H}_4$ at different structures. (b) Histogram of the CI coefficient of 20 most populated configurations for three selected structures. The inset shows the definition of structural parameters. (c, d, e) Configuration graphs corresponding to the three selected structures, respectively. The largest yellow nodes in panels (c-e) correspond to the Hartree Fock determinant.
  }
  \label{H4}
\end{figure}

We first discuss the configuration graph of a model system composed of four hydrogen atoms on the vertexes of a rectangle.
The structure is determined by two parameters, namely (i) the angle of two diagonals ($\theta$) and (ii) the length of half diagonal ($R$), as shown in the inset of Fig. \ref{H4}(b).
Via tuning the two structure parameters, we can cover a wide range of electronic structures from weak to strong correlations.
Using FCIQMC, we calculated the ground state of the rectangle $\text{H}_4$ system at various structures, \(R\) ranging from 1.5 Bohr to 8.0 Bohr and \(\theta\) ranging from \(20\degree\) to \(80 \degree\), and the energy results are shown in Fig. \ref{H4}(a).
Fig. \ref{H4}(b) is a one-dimensional plot of the 20 most populated determinant configurations of three selected structures.
Fig. \ref{H4}(c-e) shows the corresponding configuration graphs, which represent the typical graphs observed for all $\text{H}_4$ structures calculated.
Graphs of other structures are presented in Fig. S1.

Among the three typical graphs, in Fig. \ref{H4}(c) and Fig. \ref{H4}(d), the HF node is very much the single dominating configuration, with only weak edges connecting the HF node and others in both graphs.
In the graph of Fig. \ref{H4}(c), the HF determinant locates far from the other nodes, which are of much higher energy. The death rate of walkers on high energy determinants is rather high, which along with the slow spawning rate suppresses the growth of walkers on these non-HF determinants. Therefore, most walkers are trapped in the HF node and rarely spawn to the other determinants. 
In Fig. \ref{H4}(d), the HF node is close to a group of nodes even without obvious edges connected, indicating some kind of connection that enables walkers to spread between nodes. 
Such connection is stronger than Fig. \ref{H4}(c), hence some of the nodes have higher walker populations. 

In contrast,  Fig. \ref{H4}(e) is a typical strongly correlated system, featuring multiple clusters of configurations on the graph. 
The main cluster contains the HF determinant and a dozen of other determinants.
All the nodes in this main cluster are low in energy and most of the nodes have a significant number of walkers populated.
Spawning among the cluster is very frequent during FCIQMC simulation.
It is also worth noting that the cluster can simultaneously contain nodes with opposite signs.
On both sides of the main cluster are two symmetric clusters of the second lightest color and second largest size on average.
The axial symmetry of the two clusters is related to spin symmetry. Further analyses are reported in Fig. S10, which show that the nodes of the two clusters are in one-to-one correspondence, each couple has the same space orbitals but the opposite spin direction, lying roughly on the axial-symmetric position of the graph.
Another larger cluster of darker and smaller nodes lies right down below the main cluster.
These three clusters accompany the main cluster and characterize the main feature of this system.
Edges within these clusters are mostly positive and wide, walkers spawn fast inside each cluster, making the node size within each cluster approximately the same. 
Edges connecting different clusters are weaker, along which walkers spawn slower, thus the number of walkers can be prominently different between clusters.

\begin{figure}[h]
  \centering
  \includegraphics[width=16.0cm]{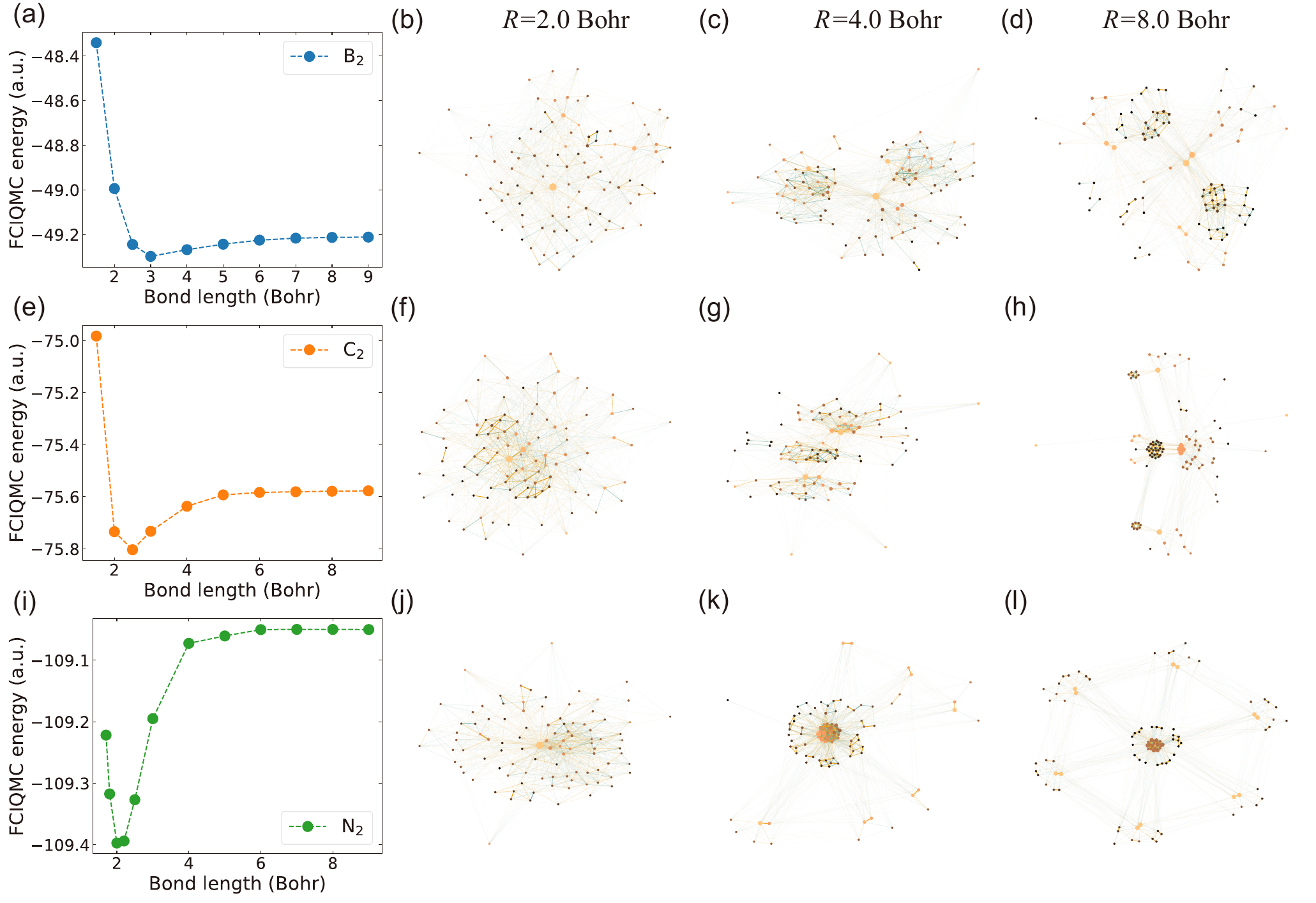}
  \caption{(a,e,i)FCIQMC energy of $\text{B}_2$, $\text{C}_2$, $\text{N}_2$. (b,c,d), (f,g,h), (j,k,l) are configuration graphs of $\text{B}_2$, $\text{C}_2$, $\text{N}_2$, respectively. The second, third, and fourth columns show configuration graphs of molecules with a bond length of 2, 4 and 8 Bohr, respectively.
  }
  \label{dimer}
\end{figure}

Having demonstrated the main characters of the designed configuration graph on the model $\text{H}_4$ system, we now discuss three realistic molecules including $\text{B}_2$, $\text{C}_2$, $\text{N}_2$. 
FCIQMC calculations of the energy of diatomic molecules have been reported in previous works as good examples to discuss the electronic correlation effects in molecules \cite{Booth_Cleland_Thom_Alavi_2011,cleland_taming_2012,jiang_general_2022}, and we only focus the graph analyses in this study. 
To highlight the change of complexity of electronic structure, we plot in Fig. \ref{dimer}(a,e,i) the energy of FCIQMC.
The energy curve shows the equilibrium bond length of the molecules and where they are to dissociate.
From the second to the fourth column of Fig. \ref{dimer}, we show three graphs for each molecule, with one structure around the equilibrium, one moderately stretched, and one strongly stretched.
More graphs at other bond lengths are enclosed in Fig. S2-S4.
In Fig. \ref{dimer}(b,j), the $\rm B_2$ and $\rm N_2$ molecules are single-reference while there are two prominent configurations in FIG. \ref{dimer}(f) for $\rm C_2$, which cannot be perceived as single-reference.
As the molecules become stretched, clustering of nodes starts to feature on the graphs.
Interestingly, the graph patterns are quite different for the three molecules, but they are common in the multi-cluster feature and suggest the systems are strongly correlated.

FCIQMC is an accurate but expensive approach because one need to converge the CI coefficient to a very high precision in order to capture the correlation energy to enough accuracy.
However, the main characters of the configuration graph can be reproduced without converging the CI coefficient to a very high precision, which will extend the use of FCIQMC in theoretical analyses of molecular electronic structure.
The construction of graph is determined by two factors: (i) the Hamiltonian matrix elements which are determined in the self-consistent-field procedure, and (ii) the configurations chosen to draw the graph. 
The information required from FCIQMC mainly comes from highly populated configurations, and those with little population would have negligible effect on the graphical features of the wavefunction. In this work, the first 100 most populated configurations are chosen to construct the graph.
The exact value of CI coefficients is to determine the size of nodes, which can be seen as secondary to our analysis when compared to the distribution of nodes and doesn't affect the character of the graph. 

\subsection{Quantification with clustering configuration descriptors}

\begin{figure}[h]
  \centering
  \includegraphics[width=16.0cm]{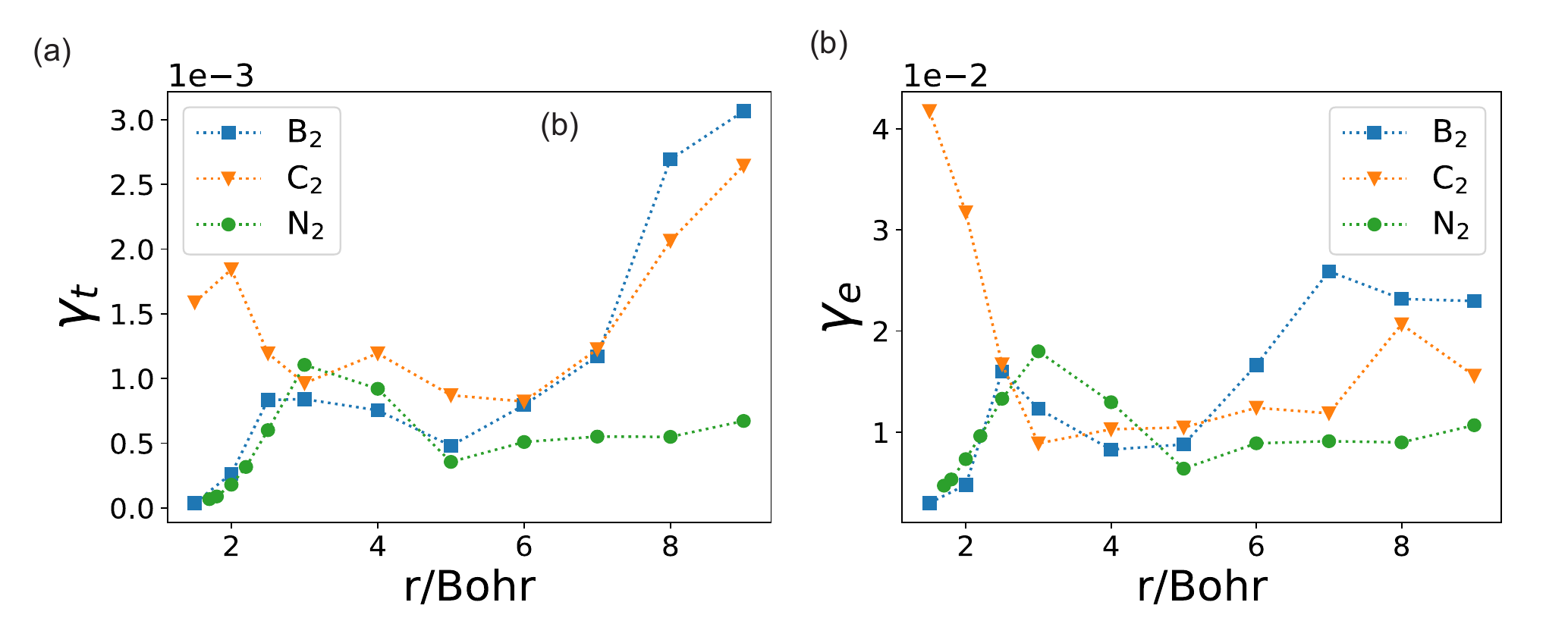}
  \caption{Descriptor $\gamma_t$ (a) and $\gamma_e$ (b) of three diatomic molecules at different bond lengths.}
  \label{descriptor}
\end{figure}

To understand the change of electronic structure in a more quantitative manner, we further analyze the multi-determinant wave functions using the two clustering configuration descriptors as defined in the Method section.
Before discussing the specific results, we would like to note that the descriptors defined are based on the insights learned from the graphic characterization. 
We aim to capture the true nature of the determinant space and the full correlation effects, including both the so-called non-dynamic and dynamic correlations.
The calculated values of the two descriptors as a function of molecular bond length of $\rm B_2$, $\rm C_2$ and $\rm N_2$ are plotted in Fig. \ref{descriptor}(a,b).
A larger value of the descriptors corresponds to a more correlated electronic structure with clusters of multiple configurations.

Overall, the two descriptors behave similarly when the bond length changes, but one can also see differences in the change of electronic structure during the dissociation of $\rm B_2$, $\rm C_2$ and $\rm N_2$.
For $\rm B_2$, the descriptors rise mostly as the bond length increases, with only a shallow decline between 3 and 5 Bohr. 
Such a trend is consistent with the conventional understanding that the ``multi-reference'' character, or in other words the ``non-dynamic correlation'', increases as the molecular bond is stretched. 
$\rm N_2$ is similar to $\rm B_2$ in that the determinant space is very much single-determinant at small bond lengths, where the descriptors gradually vanish. 
However, at large bond lengths above 4 Bohr, we do not observe the increase of descriptors, which are consistent with the qualitative visualization of configuration graphs reported in Fig. S4.
Looking closely, we can observe a weak maximum of descriptors around 3-4 Bohr, which indicates that in $\rm N_2$ the correlations are the strongest in the half-dissociation regime.
Interestingly, this observation has also been highlighted in recent neural network wave function works \cite{pfau2020ab}. 
$\rm C_2$ displays similar behavior to $\rm B_2$ at large bond lengths, but at small bond lengths its feature is opposite to $\rm B_2$ and $\rm N_2$, where the descriptor values increase during bond shrinking.
The behavior is expected from the visualization of configuration graphs presented in Fig. \ref{dimer} and Fig. S3.
This is also not of a surprise because of the known multi-reference nature of $\rm C_2$ at small bond lengths related to the existence of many low-lying states \cite{stallcop_potential_2000}.

\begin{figure}[h]
  \centering
  \includegraphics[width=16.0cm]{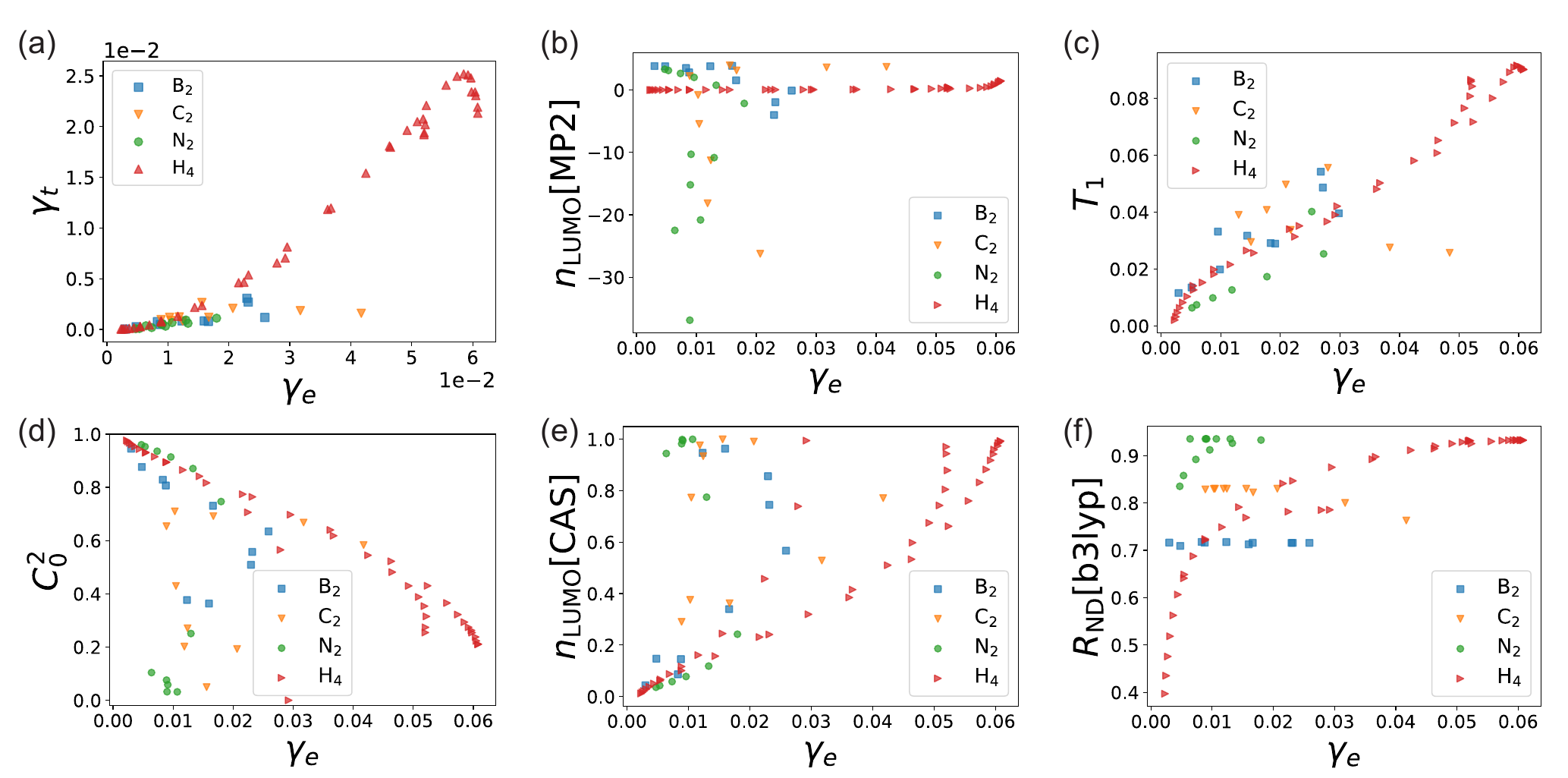}
  \caption{\textbf{The correlation between different descriptors.} (a) The correlation between $\gamma_t$ and $\gamma_e$. (b-f) The correlation between traditional descriptors and $\gamma_e$.
  }
  \label{descriptor_comparison}
\end{figure}

The quantitative results of our descriptors are consistent with the visualization of configuration graphs.
The correlation between the two descriptors is further shown in Fig. \ref{descriptor_comparison}(a), where we plot $\gamma_t$ against $\gamma_e$.
In addition to the three diatomic molecules, we also include the results on 52 $\rm H_4$ structures, extending to systems with larger descriptors.
The two outliers of $\rm C_2$ correspond to the 1.5 Bohr and 2.0 Bohr structures, where the configuration spaces contain a dominating pair of nodes.
This minor difference results in a relative underestimate of the triangular descriptor $\gamma_t$.
Apart from these special cases, the two descriptors have consistent performances.

The descriptors proposed in this work are based on the fundamental information of the determinant space, providing a measure of the full correlation effects, whereas many existing diagnostics are designed to measure the dynamic and the non-dynamic correlation separately.
Here, we further compare the other diagnostics to $\gamma_t$ and $\gamma_e$ to learn whether different approaches agree on the description of the electronic structure, in terms of whether different measures have a similar trend across different systems. 
We have calculated 12 widely used descriptors on $\text{B}_2$, $\text{C}_2$, and $\text{N}_2$.
These include three descriptors obtained from the complete active space self-consistent field (CASSCF) calculation, namely the leading weight ($C_0^2$)\cite{Lee_Taylor_2009,Sears_Sherrill_2008,Langhoff_Davidson_1974}, the occupation number of the lowest unoccupied molecular orbital ($n_{\text{LUMO}}[\text{CAS}]$) and the occupation number of the highest occupied molecular orbital ($n_{\text{HOMO}}[\text{CAS}]$).\cite{Fogueri_Kozuch_Karton_Martin_2013,Tishchenko_Zheng_Truhlar_2008}
Two similar descriptors are from the second-order M{\o}ller-Plesset perturbation theory (MP2), namely $n_{\text{LUMO}}[\text{MP2}]$ and $n_{\text{HOMO}}[\text{MP2}]$\cite{Moller_Plesset_1934}.
Three descriptors are obtained from the coupled cluster theory. $T_1$ is the Frobenius norm of the single excitation amplitude vector $t_1$\cite{Lee_Taylor_2009}; $D_1$ is the largest eigenvalue of the matrix $t_1t_1^\dagger$\cite{Janssen_Nielsen_1998}; and $D_2$ is the largest eigenvalue of the matrix $t_2t_2^\dagger$\cite{Nielsen_Janssen_1999}, where $t_2$ is the double excitation vector.
Using density functional theory (DFT) with fractional occupation, one can define the so-called dynamic ($I_{\text{D}}$) and none-dynamic ($I_{\text{ND}}$)\cite{Ramos-Cordoba_Salvador_Matito_2016,Ramos-Cordoba_Matito_2017} correlation based on occupation numbers. 
We include $I_{\text{ND}}$ and the portion of none-dynamic correlation, $r_{\text{ND}} = \frac{I_{\text{ND}}}{I_{\text{ND}} + I_{\text{D}}}$\cite{Kesharwani_Sylvetsky_Kohn_Tew_Martin_2018}, calculated using two different exchange correlation functionals.

The values of these descriptors as a function of bond length for $\rm B_2$, $\rm C_2$ and $\rm N_2$ are presented in Fig. S9.
In general, we find the descriptors based on the same type of calculation would behave similarly.
Therefore, in Fig. \ref{descriptor_comparison}(b-f), we only plot the correlation between five selected descriptors and $\gamma_e$ to avoid repetition.
The correlations to $\gamma_t$ are plotted in Fig. S12 and the quantification with correlation coefficients is plotted in Fig. S11.
Among all the descriptors considered, only coupled cluster theory based descriptors, e.g. $T_1$ in Fig. \ref{descriptor_comparison}(c), correlate relatively well with $\gamma_e$.
The others do not show a clear correlation to $\gamma_e$ and a consistent behavior for different molecules.
The poor correlations are understandable because while the traditional descriptors based on less accurate calculations can measure well the simple multi-determinant character, they may not capture all the information of the determinant space, especially the systems that present complex clustering of multiple determinants.
The same conclusion can be reached in a different test set using FCIQMC data on $\rm H_2O$ reported in a previous work \cite{jiang_general_2022}, and the analyses are presented in Fig. S13.

It is also worth noting that a descriptor is usually proposed based on a particular electronic structure theory, which may also lead to inaccuracy. 
Although the metrics used in traditional diagnostics can be suitable and reliable, the inaccuracy of the underlying method can still lead to instability when analyzing the electronic structure. The advantage of our analyses is largely due to the unbiased sampling of the determinant space by FCIQMC. Our descriptors measure the population and connections of configurations, are closely related to the intrinsic nature of the determinant space and correspond well with the visualized configuration graphs. 
In contrast, descriptors based on lower-level theories such as DFT and MP2 are affected by the inaccuracy of such calculations, in particular on systems with strong correlations.
Therefore, these descriptors may only work well for some systems, but the performances usually vary from one to another and become increasingly unreliable for strongly correlated systems.
In principle, our descriptors can apply to other multi-determinant wave functions, but the performance would depend on the quality of the underlying calculations. 
Hence, if accurate benchmark wave functions are available, the descriptors can also be used to measure the accuracy of an approximated wave function calculation.
For other diagnostics, improvement in performance can also be expected when the corresponding method is further developed, e.g. employing a more accurate DFT functional.

\subsection{Discussion}

\begin{figure}[h]
  \centering
  \includegraphics[width=16.0cm]{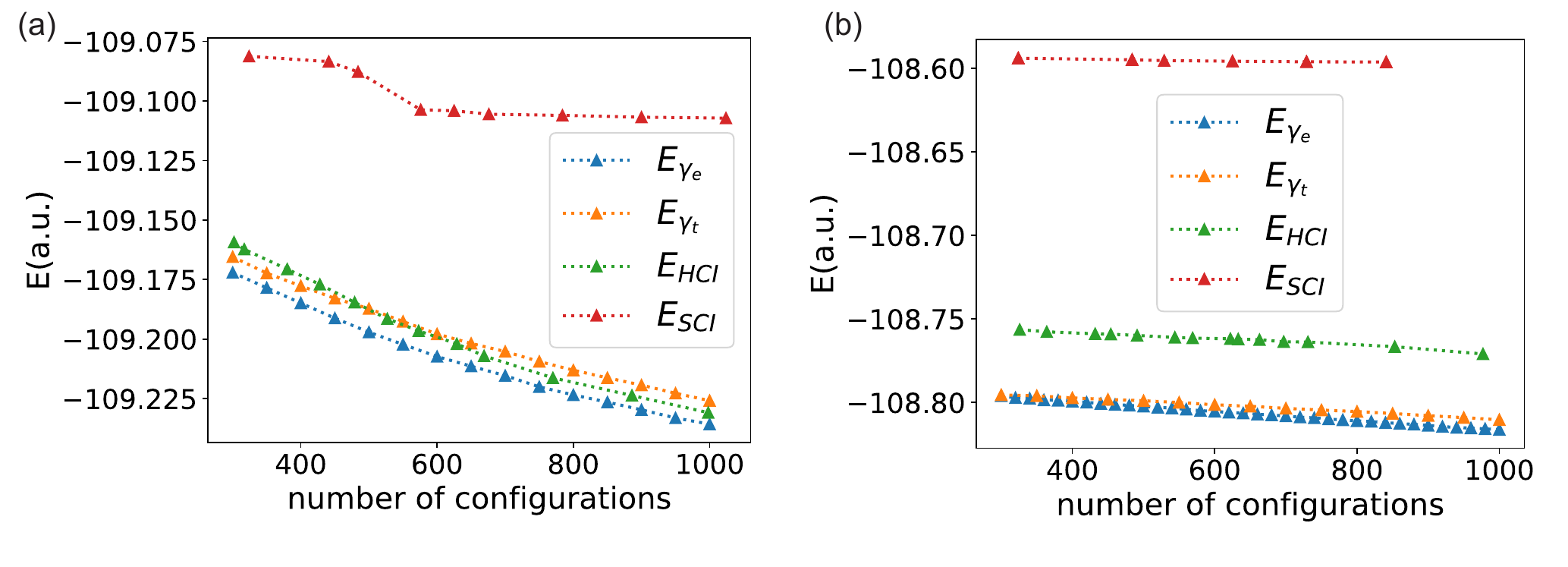}
  \caption{CI energy as a function of the number of configurations computed using different selections of configurations. 
  Panels (a-b) are calculations of $\rm N_2$ with bond length at 2 Bohr and 7 Bohr, respectively. Blue and orange curves result from a direct selection of configurations based on $\gamma_e^u$ and $\gamma_t^u$, respectively.
  Green and red curves are calculated using heat-bath CI (HCI) and selected CI (SCI) as implemented in PySCF. The dependence of $\rm {E}_{HCI}$ and $\rm {E}_{SCI}$ on the number of configurations are obtained via changing their cutoff thresholds.}
  \label{fig:myci}
\end{figure}

Before concluding, we would like to point out that the insights provided by these analyses can assist further developments of CI-based wave function methods, especially those aimed at tackling the most challenging strongly correlated system. 
In existing methods such as selected CI \cite{ivanic_identification_2001} and heat-bath CI \cite{holmes_heat-bath_2016}, configurations are selected based on a single threshold. Our clustering configuration descriptors provide an alternative option for selecting configurations, so that the calculations can achieve high accuracy with fewer configurations included. To demonstrate the possible usefulness of our descriptors, we plot the energy computed as a function of the number of configurations based on different selection schemes in Fig. \ref{fig:myci}. $\rm {E}_{\gamma_e}$, $\rm {E}_{\gamma_t}$, $\rm {E}_{HCI}$ and $\rm {E}_{SCI}$ are energies obtained following selection rules based on the $\gamma_e^u$, $\gamma_t^u$, the heat-bath CI (HCI) rule and the selected CI (SCI) rule, respectively. 
$\gamma_e^u=\sum_{v \ne u} (C_u)^2 (C_v)^2 |H_{uv}| $ describes the contribution of configuration $u$ to the descriptor, namely $ \gamma_e = \sum_u \gamma_e^u $. 
Similarly, we can define $\gamma_t^u$, which gives $ \gamma_t = \sum_u \gamma_t^u $.
The results of SCI and HCI with different numbers of configurations are obtained by changing the cutoff threshold, larger cutoffs lead to fewer selected configurations.

We see that selecting by $\gamma_e^u$ gives the lowest energy for the same number of configurations. 
Selecting by $\gamma_t^u$ also works well, which is resulted from the consistency of the two descriptors.
Panels (a) and (b) show the results on two different systems. The advantage is more significant for systems with larger $\gamma_e$ (stretched $\rm N_2$ in panel b) than systems with smaller $\gamma_e$ (short $\rm N_2$ in panel a). 
As shown in Fig. S14, the effectiveness of $\gamma_e$ based selection scheme also applies to the $\rm H_4$ system, where it is also clear that the selection follows the clustering feature of the configuration graph.
The results suggest that the clustering configuration descriptors are likely to be helpful for further development of existing methods, such as selected CI \cite{holmes_heat-bath_2016}, heat-bath CI \cite{ivanic_identification_2001}, iterative CI \cite{liu_ici_2016}, etc. In addition to the direct use of clustering configuration descriptors, there are several other options indicated by the graph analysis. For example, one can select a whole cluster of configurations to perform CI calculations; one can apply existing configuration selection methods within clusters identified; one can utilize the graphical symmetry observed when considering the configuration selections.
These options are to be implemented and tested in the future.

\section{Conclusions}

In this study, we propose new analyses for the multi-determinant wave function of molecules.
The analyses include (i) a graphic representation of the populations and connections between important configurations of the determinant space; (ii) two analytic descriptors for quantitative characterization of the determinant space.
In conjunction with the unbiased sampling of FCIQMC, our characterizations provide a more complete description of the true nature of the electronic structure of molecules, complementing the existing characterizations and diagnostics.
The possibilities of further development of multi-determinant wave function methods using the characterizations presented are also discussed.

\section{acknowledgement}
This work was supported by the National Natural Science Foundation of China under Grant No. 92165101 and No. 11974024, the National Key R\&D Program of China under Grant No. 2021YFA1400500, and the
Strategic Priority Research Program of Chinese Academy of Sciences under Grant No. XDB33000000.
We are grateful for computational resources provided by Peking University, the TianHe-1A supercomputer, Shanghai Supercomputer Center, and Songshan Lake Materials Lab.

\bibliography{ref.bib}

\end{document}